\documentclass[12pt]{article}
\usepackage{graphics, color}
\usepackage{amssymb}
\usepackage{amsmath}
\usepackage{feynmf}
\begin{document}

\baselineskip=15.5pt
\pagestyle{plain}
\setcounter{page}{1}

%Title page

\begin{titlepage}
\bigskip
\rightline{hep-th/0605028}
\bigskip\bigskip\bigskip
\centerline{\Large \bf The $SU(2)$ Long Range Bethe Ansatz}
\medskip
\centerline{\Large \bf and Continuous Integrable Systems}
\bigskip\bigskip\bigskip

\centerline{\large Nelia Mann$^a$}
\bigskip
\centerline{\em $^a$ Department of Physics, University of California,} \centerline{\em Santa Barbara, CA, 93106} \centerline{\em nelia@physics.ucsb.edu}
\smallskip
\bigskip
\bigskip
\bigskip\bigskip

%ABSTRACT

\begin{abstract}
We explore the relationship between the $SU(2)$ sector of a general integrable field theory and the all-loop guess for the anomalous dimensions of $SU(2)$ operators in $\mathcal{N} = 4$ super Yang-Mills theory.  We demonstrate that the $SU(2)$ structure of a nested Bethe ansatz alone reproduces much of the all-loop guess without depending on the details of the particular field theory.  We speculate on the implications of this for strings in AdS$_{5} \times$S$^5$ being described by the multi-particle states of an integrable worldsheet theory, and relate the techniques here to the known relationship between the Hubbard model and the all-loop guess.
\medskip
\noindent
\end{abstract}
\end{titlepage}

\section{Introduction}

The possible correspondence between type IIB string theory in an AdS$_{5} \times$S$^5$ background and $\mathcal{N} = 4$ super Yang-Mills theory has been one of the most interesting areas of study in string theory since its initial proposal by Maldacena \cite{Maldacena}.  Ideally, we would like to demonstrate this correspondence by comparing the energies of string states to the dimensions of operators in the field theory.  Until quite recently this was only possible for a small selection of states highly protected by supersymmetry.  However, it is now understood that the techniques of integrability can be used to compare larger sectors of the theories and may eventually demonstrate complete correspondence between the two theories in the strict large-N limit, where gauge theory calculations come entirely from planar graphs and where the string-theory is non-interacting.  

On the gauge theory side, Minahan and Zarembo initially demonstrated that the one-loop dilatation operator acting on scalar, single-trace operators could be completely diagonalized with a Bethe ansatz by interpreting the single-trace operators are periodic spin chains \cite{MZ}.  This was quickly generalized to fully supersymmetric $PSU(2,2|4)$ spin chains \cite{BS1}.  Extension to two-loops followed, along with extension in some sectors to three-loops \cite{BKS, su23}.  So far, all gauge theory calculations support integrability surviving to all loops.  In addition, the gauge theory calculations known seem to obey ``BMN scaling''.  If both BMN scaling and integrability are satisfied to all loops, then these conditions together generate a guess for the anomalous dimensions of operators to all loops \cite{BDipS, BS05}.  Now, this guess is not believed to be correct beyond $J$ loops for an operator of bare dimension $J$ so that the all loop exact expression is really only applicable to infinitely long operators.

On the string side, research into integrability started with the discovery of an infinite number of Yangian symmetries in the classical AdS$_{5} \times$S$^5$ string theory \cite{BPR}, which indicate integrability at least at the classical level.  In addition, various semi-classical string states have been constructed \cite{GKP, FT, FT2} and compared to the long operators in the gauge theory \cite{BMSZ, BFST}.  In the strict BMN limit the energies and dimensions agree, though away from this disagreement has been found at three loop order that can probably be attributed to a difference in the order or limits taken on either side of the correspondence \cite{SS}.  The classical charges have been used to generalize this and encode in integral equations all possible semi-classical string states and their energies \cite{KMMZ, BKSZ}, which can be compared to similar integral equations on the gauge side \cite{S-N, BKSZ2}.  Finally, work has been done into how a Bethe ansatz can be applied to the string theory as a 1+1 dimensional worldsheet theory in the light-cone gauge \cite{AFS} - \cite{LC4} and without this gauge fixing \cite{MP, MP2, Kazakov, Zarembo}.  These techniques assume infinitely long strings just as the gauge theory calculations assume infinitely long operators.

The Bethe Ansaetze are different on the two sides of the correspondence because of the differences between spin chain and continuous systems.  In the gauge theory, at each loop order there is a spin chain Hamiltonian.  At one loop, the Hamiltonian has only nearest-neighbor interactions at two loops, the interactions extend to next-nearest-neighbor, and so on.  At any given loop this generates a discrete short-range interaction, but apparently as we extend to include all loops we can reproduce the smoothed-out continuous interaction present in the 1+1 dimensional worldsheet theory.  In this paper, we study how the correspondence between an all-loop spin chain and a continuously integrable structure occurs using an $SU(2)$ sector on both sides.  Specifically, we will demonstrate where a loop-like expansion from the all-loop guess can come from in a continuously integrable system.

In section two we will review the $SU(2)$ sector all-loop Bethe ansatz from the gauge theory and an $SU(2)$ sector in a continuously integable system, and discuss some clues as to where the gauge theory loop expansion will come from on the string side.  In section three we will use these clues to produce a loop-like expansion structure from the continuous Bethe ansatz.  We will discuss how this expansion works in a conformal 1+1 dimensional theory obtained as the limit of a massive theory, as well as how this expansion relates to the proposed correspondence between the all-loop guess and the integrable Hubbard model of electrons on a lattice \cite{Hubbard1, Hubbard2, Hubbard3}.  In section four we will conclude with the implications of this work as well as some open questions that remain to be answered.

\section{Short Review of Continuous and Spin Chain Integrability in AdS/CFT}

In order to create a comparison between the spin chain and continuous systems, we first need to review the known results for Bethe ansaetze on both sides of the correspondence.  Below we review the spin chain Bethe ansatz structure used in the gauge theory and present the all-loop ansatz.  Then we will discuss the $SU(2)$ sector in a continuously integrable system, and how the Virasoro constraints of a string theory might be applied here.  Finally we will discuss the supercoset theory $OSP(2n+2|2n)$ as a toy model for the worldsheet theory and explore the clues it gives us as to where the loop expansion of the dual gauge theory will arise.

\subsection{Review of the SU(2) Long Range Bethe Ansatz}

It has long been known now that an $SU(2)$ sector of scalar operators in planar $\mathcal{N} = 4$ SYM can be studied through the use of spin chains \cite{MZ}.  In particular, the anomalous dimension of an operator of the form 
\begin{eqnarray}
\mathcal{O} = \mbox{Tr} X^{J_{2}}Z^{J_{1}} + \cdots
\end{eqnarray}
can be seen as the energy of an $SU(2)$ spin chain of length $J = J_{1} + J_{2}$, where the Bethe reference state is the operator
\begin{eqnarray}
\mathcal{O} = \mbox{Tr} Z^{J}
\end{eqnarray}
which is BPS and known to have vanishing anomalous dimension.  The anomalous dimension can be calculated in the gauge theory as a loop expansion, and it has been shown that the one-loop dilitation operator forms the (integrable) $XXX$ nearest-neighbor Hamiltonian on the $SU(2)$ spin chain.  The one-loop anomalous dimension can then be calculated using a Bethe Ansatz to be
\begin{eqnarray}
\gamma = \frac{\lambda}{2\pi^2}\sum_{\alpha = 1}^{J_{2}} \sin^2 \frac{p_{\alpha}}{2}
\end{eqnarray}
\begin{eqnarray}
\left(\frac{u_{\alpha} + i/2}{u_{\alpha} - i/2}\right)^{J} = \prod_{\beta \ne \alpha}^{J_{2}} \frac{u_{\alpha} - u_{\beta} + i}{u_{\alpha} - u_{\beta} - i} \equiv e^{ip_{\alpha}J}
\end{eqnarray}
The spectral parameters $u_{\alpha}$ take quantized values in the complex plane because of the second equation, as we take $J$ and $J_{2}$ large these values condense into continuous cuts in the complex plane labeled by mode numbers.

If we expand the dilatation operator beyond one loop, we find that the associated spin chain obtains longer range interactions: at two loops we have next-nearest-neighbor interactions, at three loops next-next-nearest-neighbor, and so forth.  The dilatation operator has been calculated in this sector up to three loops and demonstrated to be integrable \cite{BKS, su23}.   Both the anomalous dimension's dependence on $\sin^2\frac{p_{\alpha}}{2}$ and the relationship between $p_{\alpha}$ and the spectral parameter $u_{\alpha}$ obtain loop corrections.  Finally, it is known \cite{BDipS, BS05}  that if we posit BMN scaling and integrability for higher loops, we can extend this to an all-loop guess that takes the relatively simple form
\begin{eqnarray}
\label{eqn8}
\gamma = \sum_{\alpha = 1}^{J_{2}} \sqrt{1 + \frac{\lambda}{\pi^2}\sin^2\frac{p_{\alpha}}{2}} - 1
\end{eqnarray}
\begin{eqnarray}
\label{eqn9}
u_{\alpha}(p_{\alpha}) = \frac{1}{2}\cot\frac{p_{\alpha}}{2}\sqrt{1 + \frac{\lambda}{\pi^2}\sin^2\frac{p_{\alpha}}{2}}
\end{eqnarray} 
\begin{eqnarray}
\label{eqn10}
e^{ip_{\alpha}J} = \prod_{\beta \ne \alpha}^{\infty} \frac{u_{\alpha} - u_{\beta} + i}{u_{\alpha} - u_{\beta} +i}
\end{eqnarray}
For a spin chain of finite length $J$, this guess is believed to break down at order $\lambda^{J}$.  However, for the duration of this paper we will be assuming a limit $J \rightarrow \infty$ where this guess should be valid to all orders.

\subsection{Review of Continuously Integrable Structure}

Continuously integrable field theories differ from integrable spin chains in that in general an integrable field theory starts from a Bethe reference state that is an empty, chargeless vacuum, while the spin chain generally starts from a reference state that has maximal physical charge (other reference states are possible, but usually much more complicated).  Because of this, the $SU(2)$ sector of an integrable field theory is built up using a two-stage, ``nested'' Bethe ansatz:
\begin{eqnarray}
e^{ip_j L} = \prod_{\beta} \frac{i\theta_j - i\Lambda_{\beta} + \frac{\pi}{C_{g}}}{i\theta_j - i\Lambda_{\beta} - \frac{\pi}{C_{g}}} \prod_{i \ne j} S_{pp}(\theta_i - \theta_j) + \mbox{casimir effects}
\end{eqnarray}
\begin{eqnarray}
\prod_{j} \frac{i\theta_j - i\Lambda_{\alpha} + \frac{\pi}{C_{g}}}{i\theta_j - i\Lambda_{\alpha} - \frac{\pi}{C_{g}}} = \prod_{\beta \neq \alpha} \frac{i\Lambda_{\alpha} - i\Lambda_{\beta} - \frac{2\pi}{C_{g}}}{i\Lambda_{\alpha} - i\Lambda_{\beta} + \frac{2\pi}{C_{g}}}.
\end{eqnarray}
Here, we are exciting $J$ 2-d particles which have rapidities $\theta_{i}$ and momenta $p_{i} = m\sinh\theta_{i}$.  $S_{pp}(\theta_{i} - \theta_{j})$ is the identical particle S-matrix element for the theory.  The particles then form a nested spin chain, where each site can be ``up'' or ``down'' in the $SU(2)$ sense.  In this spin chain, we start by assuming that all particles are the same ($J_{2} = 0$) and then place impurities in the spin chain.  These impurities then have ``pseudorapidities'' $\Lambda_{\alpha}$.  $C_{g}$ is a group-theoretic factor that is determined by the specifics of the theory our $SU(2)$ sector originates from.  The $SU(2)$ structure determines everything but the identical particle S-matrix and this number.  The energy of a state determined by these equations is then
\begin{eqnarray}
E = \sum_{i = 1}^{J} m\sinh\theta_{i}.
\end{eqnarray}
Suppose that we are looking at a worldsheet theory that is meant to produce strings dual to the operators we discussed in the gauge theory.  Here there are $J + J_{2}$ degrees of freedom (the $\theta_{i}$s and the $\Lambda_{\alpha}$s) while there were only $J_{2}$ degrees of freedom on the gauge side.  On the other hand, this structure does not take into account Virasoro constraints that would have to be imposed in a string theory.  It has been proposed that the effect of these constraints would be to completely determine the $\theta_{i}$s in terms of the $\Lambda_{\alpha}$s, and specifically to require that the quantization of the $\theta_{i}$s all have the same mode number \cite{MP2, Kazakov}.  If we are interested in comparing with operators that are infinitely long, we can consider the limit $J \rightarrow \infty$, $L \rightarrow \infty$ such that $j = J/L = \mbox{fixed}$, and the top equation becomes an integral equation for the density of particles in rapidity space $\rho(\theta) = \frac{1}{L(\theta_{i+1} - \theta_{i})}$.  The mode number constraint on the $\theta$s then requires that they fall into a single, continuous range which we demand to be symmetric $-B < \theta < B$ so that the worldsheet momentum vanishes.  Notice that in this limit the Casimir effects vanish so that we have exact expressions.  The equations then become
\begin{eqnarray}
\label{eqn1}
\rho(\theta) = \int_{-B}^{B} K(\theta - \theta')\rho(\theta') \ d\theta = \frac{m}{2\pi}\cosh\theta - \frac{1}{C_{g}L}\sum_{\alpha = 1}^{J_{2}} \frac{1}{\frac{\pi^2}{C_{g}^2} + (\Lambda_{\alpha} - \theta)^2}
\end{eqnarray}
\begin{eqnarray}
\label{eqn2}
\prod_{i = 1}^{J} \frac{i\theta_{i} - i\Lambda_{\alpha} + \frac{\pi}{C_{g}}}{i\theta_{i} - i\Lambda_{\alpha} - \frac{\pi}{C_{g}}} = \prod_{\beta \ne \alpha} \frac{i\Lambda_{\alpha} - i\Lambda_{\beta} - \frac{2\pi}{C_{g}}}{i\Lambda_{\alpha} - i\Lambda_{\beta} + \frac{2\pi}{C_{g}}}
\end{eqnarray}
where $K(\theta) = \frac{1}{2\pi i}\frac{d}{d\theta}\ln S(\theta)$.  The second equation could also be turned into an integral equation in this limit, but for now we will leave it as is.  Since the AdS$_{5} \times$S$^5$ sigma model should be conformal, it is expected that to study it using a Bethe ansatz we would need to find some massive deformation of this theory and then take a conformal limit, which we will discuss in the next section.  The $\Lambda_{\alpha}$s here take quantized values in the complex plane, and condense into continuous cuts labeled by mode numbers, just like the $u_{\alpha}$s.  

\subsection{Clues from the $OSP(2n+2|2n)$ toy model}

Now, the exact S-matrix for the AdS$_{5} \times$ S$^5$ sigma model is not known, but various toy models for the worldsheet theory have been studied in the Bethe ansatz context for which the exact S-matrix can be written down \cite{MP2, Kazakov, Zarembo}.  Looking at these toy models can give us insight into the problem of relating the continuous integrability story to the spin chain story.  In particular the $OSP(2n + 2|2n)$ conformal sigma model studied in \cite{MP2}, which has an $O(6)$ sector that is classically identical to the $S^5$ part of the AdS$_{5} \times$ S$^5$ theory is a valuable test case for taking a conformal limit.  It was shown that we can apply a Bethe ansatz to this theory by considering the limit of $OSP(2n+N|2n)$ as $N \rightarrow 2$ and $m \rightarrow 0$.  In this case we have $C_{g} = N - 2$ and the integral equation determining $\rho(\theta)$ has a well-defined conformal limit where it splits up into three equations: one for right-movers, one for left-movers, and one for particles which have zero momentum in the limit.  Dependence on a dimensionless coupling is recovered through a factor $\chi = (N - 2)\ln \frac{\mu}{m}$ which is kept fixed in the limit, and in the classical, large $\chi$ limit of the theory we have $\chi \approx \lambda$.

If we consider a state where $J_{2}$ gets large with $J$ and $L$, in the large $\chi$ limit we can reproduce the ``spinnning string'' solutions of semi-classical AdS$_{5} \times$S$^5$.  These solutions are known to be related to operators in the field theory given by specific configurations of cuts in the complex plane of condensed $u_{\alpha}$s, and here the configurations of $\Lambda_{\alpha}$s match those of the $u_{\alpha}$s.  Thus we suppose that these represent the same degrees of freedom, with some rescaling.

We also know from studying states where the number of impurities is finite that in these cases the impurities behave like massive particles.  The momenta of these impurity particles is determined by the associated $\Lambda_{\alpha}$, and large $\Lambda_{\alpha}/\chi$ corresponds to small momenta-- non-relativistic motion.  If we make an approximation that $\Lambda_{\alpha}/\chi$ is large, then in the classical limit relativistic corrections come in the form 
\begin{eqnarray}
\frac{v^2}{c^2} = \frac{n\lambda}{J^2}
\end{eqnarray}
where $n$ is the mode number associated to $\Lambda_{\alpha}$ and is small for $\Lambda_{\alpha}$ large.   In the semi-classical limit, this expansion fails to exactly reproduce the gauge theory results starting at the three-loop order.  However, perhaps if we do not work in the semi-classical limit but rather in a coupling regime where $\lambda$ is small, an expansion in relativistic corrections on the string side might reproduce a loop-like expansion.

It is important to stress that ``spinning string'' states receive several types of corrections.  First, they receive finite size corrections in $1/L$, and they receive quantum corrections like $1/\sqrt{\lambda}$, and finally they receive relativistic corrections like $n\lambda/J^2$.  In the large $L$ large $\lambda$ limit, we can still write down equations that are exact in relativistic corrections; here the Bethe ansatz equations reproduce the KMMZ equation for semi-classical spinning strings.  This equation is known to disagree with the gauge theory all-loop guess at three loops.  However, it seems clear that it should receive both finite size and quantum corrections, while the gauge theory guess will certainly receive finite size corrections but might otherwise be correct.

\section{The Loop-like Expansion in the Worldsheet Picture}

Now, we would like to consider an $SU(2)$ sector in a general integrable 1+1 dimensional field theory.  We won't assume any knowledge about the details of the theory beyond the $SU(2)$ structure, this way we can assume that the theory is, in fact, the AdS$_{5} \times$S$^5$ worldsheet theory, without knowing much about the theory.  We want to exploit the structure of the $SU(2)$ nested Bethe ansatz to produce a ``relativistic'' expansion, which we believe will have something to do with the loop expansion of the anomalous dimension in the dual theory.

Let us consider first equation (\ref{eqn1}), which can be rewritten as
\begin{eqnarray}
\rho(\theta) = \int_{-B}^{B} K(\theta - \theta')\rho(\theta') \ d\theta = \frac{m}{2\pi}\cosh\theta - \frac{C_{g}}{4\pi^2L}\sum_{\alpha = 1}^{J_{2}} \frac{1}{\frac{1}{4} + (v_{\alpha} - \phi)^2}
\end{eqnarray}
where $v_{\alpha} = \frac{C_{g}\Lambda_{\alpha}}{2\pi}$ and $\phi = \frac{C_{g}\theta}{2\pi}$.  Notice that, becuase we are assuming a symmetric range of rapidities, only the symmetric part of source terms for this integral equation influence the energy of the system.  Following our inspiration to assume that a loop-like expansion will be an expansion in small $\phi$ as compared to the $v_{\alpha}$, we take the second source term and write it as an infinite sum, using the rule
\begin{eqnarray}
\label{expand}
\mbox{Even}\left(\frac{1}{\frac{1}{4} + (v - \phi)^2}\right) = \sum_{n = 0}^{\infty} \phi^{2n}Q_{n}\left(\frac{1}{v^2 + \frac{1}{4}}\right)
\end{eqnarray}
with the $Q_{n}$ polynomials defined by
\begin{eqnarray}
Q_{n}(x) = x^{n+1}\sum_{k = 0}^{n} \frac{2n+1}{2k+1}\left(\begin{array}{c} n+k \\ 2k \end{array}\right) (-x)^{k} = x^{n+1}\left[U_{n}\left(1 - \frac{x}{2}\right) + U_{n-1}\left(1 - \frac{x}{2}\right)\right]
\end{eqnarray}
where the $U_{n}(x)$ are Chebyshev polynomials of the second kind.\footnote{The relationship of the $Q_{n}(x)$ to the Chebyshev polynomials was demonstrated to me by F. Cachazo.}  Now, because the integral equation is linear in the source terms, we can write the density $\rho(\theta)$ as an infinite sum
\begin{eqnarray}
\rho(\theta) = \rho_{pure}(\theta) + \rho_{0}(\theta) + \rho_{1}(\theta) + \rho_{2}(\theta) + \cdots
\end{eqnarray}
with
\begin{eqnarray}
\rho_{pure}(\theta) - \int_{-B}^{B} K(\theta - \theta')\rho_{pure}(\theta') \ d\theta' = m\cosh\theta
\end{eqnarray}
and
\begin{eqnarray}
\rho_{n}(\theta) - \int_{-B}^{B} K(\theta - \theta') \rho_{n}(\theta') \ d\theta' = -\phi^{2n}\frac{1}{2\pi L}\sum_{\alpha = 1}^{J_{2}} Q_{n}\left(\frac{1}{v_{\alpha}^2 + \frac{1}{4}}\right).
\end{eqnarray}
By this technique, we effectively decouple the $\rho(\theta)$ integral equation from dependence on the $v_{\alpha}$, so that the only dependence of the $\rho_{n}(\theta)$ on the $v_{\alpha}$ is given by an overall factor for each function $\rho_{n}(\theta)$.  This is then transfered to the energy of the state, which is then written as an infinite sum using these polynomials:
\begin{eqnarray}
E = E_{pure} + \sum_{n = 0}^{\infty} g_{n}(\lambda)Q_{n}\left(\frac{1}{v_{\alpha}^2 + \frac{1}{4}}\right)
\end{eqnarray}
where the unknown constants $g_{n}$ depend on the solutions to integral equations; they are independent of the $v_{\alpha}$ but presumably depend on the coupling constant $\lambda$.  This energy should be related to the dimensions of operators.  This could then also be expanded in terms of these polynomials.

We can compare this with the loop expansion of the gauge theory anomalous dimension.  Writing equation (\ref{eqn8}) in terms of the spectral parameters using equation (\ref{eqn9}) we get
\begin{eqnarray}
\gamma = \sum_{\alpha = 1}^{J_{2}} \sqrt{1 + \frac{\lambda}{2\pi^2} - 2\left(\frac{1}{4} + u_{\alpha}^2\right) + \sqrt{\left(2\left(\frac{1}{4} + u_{\alpha}^2\right) - \frac{\lambda}{2\pi^2}\right)^2 + \frac{\lambda}{\pi^2}}} - 1.
\end{eqnarray}
We find\footnote{This can most easily be demonstrated by relating the double square root expression to the function in (\ref{expand}) through a contour integral; this method was shown to me by T. Erler.} that an expansion in $\lambda$ gives
\begin{eqnarray}
\label{eqn3}
\gamma = \sum_{n = 0}^{\infty} \frac{2(2n)!}{(n+1)!n!}\left(\frac{\lambda}{16\pi^2}\right)^{n+1} \left[\sum_{\alpha = 1}^{J_{2}} Q_{n}\left(\frac{1}{\frac{1}{4} + u_{\alpha}^2}\right)\right]
\end{eqnarray}
where the $Q_{n}$ are the same polynomials that appeared before.  The appearance of these polynomials in both places is strongly suggestive that we should have $u_{\alpha} = v_{\alpha}$ and that this relativistic expansion is exactly the loop expansion of the gauge theory.

The other aspect of relating worldsheet Bethe equations to spin chain equations is to relate the equations quantizing the $v_{\alpha}$ (\ref{eqn2}) to those quantizing the $u_{\alpha}$, (\ref{eqn9}) and (\ref{eqn10}).  We rewrite (\ref{eqn2}) as
\begin{eqnarray}
e^{ip_{\alpha}J} \equiv \prod_{i = 1}^{J} \frac{v_{\alpha} - \frac{C_{g}\theta_{i}}{2\pi} + \frac{i}{2}}{v_{\alpha} - \frac{C_{g}\theta_{i}}{2\pi} - \frac{i}{2}} = \prod_{\beta \ne \alpha} \frac{v_{\alpha} - v_{\beta} + i}{v_{\alpha} - v_{\beta} - i}
\end{eqnarray}
and using the fact that $\sum_{i = 1}^{J} \theta_{i}^{2n + 1} = 0$, (because the $\theta_{i}$s are arranged symmetrically) we can expand $p_{\alpha}$ in relativistic corrections, obtaining
\begin{eqnarray}
\label{eqn4}
\frac{p_{\alpha}}{2} = \sin^{-1}\sqrt{\frac{1}{1 + 4v_{\alpha}^2}} + \frac{v_{\alpha}}{2}\sum_{n = 0}^{\infty}h_{n+1}(\lambda)P_{n}\left(\frac{1}{\frac{1}{4} + v_{\alpha}^2}\right)
\end{eqnarray}
with
\begin{eqnarray}
P_{n}(x) = \frac{x^{n+2}}{n+1}\sum_{\ell = 0}^{n} \left(\begin{array}{c} n + \ell + 1 \\ 2\ell + 1 \end{array}\right) (-x)^{\ell} = \frac{x^{n+2}}{n+1}U_{n}\left(1 - \frac{x}{2}\right)
\end{eqnarray}
and
\begin{eqnarray}
h_{n}(\lambda) = \frac{1}{J} \sum_{i = 1}^{J} \left(\frac{C_{g}\theta_{i}}{2\pi}\right)^{2n} = \frac{1}{j}\int_{-B}^{B} \left(\frac{C_{g}\theta}{2\pi}\right)^{2n}\rho(\theta) \ d\theta
\end{eqnarray} 
where, again, the $U_{n}(x)$ are Chebychev polynomials of the second kind.

We can compare this with the loop expansion from the gauge theory where we have
\begin{eqnarray}
\frac{p_{\alpha}}{2} = \sin^{-1}\sqrt{\frac{\frac{\lambda}{4\pi^2} - \left(\frac{1}{4} + u_{\alpha}^2\right) + \sqrt{\left(\left(\frac{1}{4} + u_{\alpha}^2\right) - \frac{\lambda}{4\pi^2}\right)^2 + \frac{\lambda}{4\pi^2}}}{\frac{\lambda}{2\pi^2}}} 
\end{eqnarray}
and then we write the expansion
 \begin{eqnarray}
 \label{eqn7}
\frac{p_{\alpha}}{2} =  \sin^{-1}\sqrt{\frac{1}{1 + 4u_{\alpha}^2}} + u_{\alpha}\sum_{n = 0}^{\infty} \left(\frac{\lambda}{16\pi^2}\right)^{n+1}\left(\begin{array}{c} 2n + 1 \\ n \end{array}\right) P_{n}\left(\frac{1}{\frac{1}{4} + u_{\alpha}^2}\right)
\end{eqnarray}
and again the polynomials appearing here are the same as those used before.  Thus we will have agreement between the two sides providing we have
\begin{eqnarray}
h_{n}(\lambda) = 2\left(\begin{array}{c} 2n - 1 \\ n \end{array}\right)\left(\frac{\lambda}{16\pi^2}\right)^{n}.
\end{eqnarray}

Now, from what is written it is not obvious that the functions $h_{n}(\lambda)$ obtained in the worldsheet theory will actually be independent of the $v_{\alpha}$.  In fact, it seems that for a general theory we could expand the $\rho(\theta)$ as before and obtain $h_{n}(\lambda)$ that depend on factors like $\sum_{\alpha = 1}^{J_{2}} Q_{m}\left(\frac{1}{\frac{1}{4} + v_{\alpha}^2}\right)$.  There is no room for terms like this if we are to match the worldsheet theory to the all loop guess; an important part of the all loop guess is that $p_{\alpha}$ should depend only on the associated $u_{\alpha}$, and not on $u_{\beta}$ for $\beta \ne \alpha$.  Because the $P_{n}(x)$ are linearly independent, the only way for this to happen is for the dependence of $h_{n}(\lambda)$ on these terms independently to disappear.  How this can happen will be discussed later.

It is important to note that the only thing being assumed here is the $SU(2)$ structure of the sector we are studying.  The expansions used will always give the right polynomial dependence as to agree with the loop expansion of the all-loop guess, and since the polynomials are linearly independent, it seems that this must be the only way to match the loop expansion.   

\subsection{A Conformal Limit of the Continuous System}

In the $OSP(2n+2|2n)$ toy model, the conformal Bethe ansatz was found by starting with a non-conformal theory, $OSP(2n+N|2n)$, and taking the $N \rightarrow 2$ limit in the Bethe equations (where the S-matrix depends explicitly on $N$) at the same time as the $m \rightarrow 0$ limit, while keeping the parameter $\chi = (N - 2)\ln \frac{\mu}{m}$ fixed.  If we suppose that the worldsheet theory for AdS$_{5} \times$S$^5$ (which is also conformal) can be studied in the same way, then we can go a little farther with the calculations of the previous section.  Now, the constant $C_{g}$ from equations (\ref{eqn1}) and (\ref{eqn2}) is a group theoretic factor that we expect to approach zero in the conformal limit, and we'll take the limit so that
\begin{eqnarray}
\chi = \frac{C_{g}}{2\pi}\ln \frac{\mu}{m}
\end{eqnarray}
stays fixed.

In this case, the worldsheet energy density is given in terms of the density of left- and right-movers as
\begin{eqnarray}
\mathcal{E} \equiv \frac{E}{L} = \frac{(1 - \tilde{K})\pi}{2}\left(j_{R} + j_{L}\right)^2
\end{eqnarray}
where
\begin{eqnarray}
\tilde{K} = \frac{1}{2\pi i}\lim_{\theta \rightarrow \infty} \left[\lim_{C_{g} \rightarrow 0} \ln \frac{S(\theta)}{S(-\theta)}\right].
\end{eqnarray}
The number densities of left- and right-movers are not good quantum numbers, so we want to express the energy density in terms of the total number density $j$, which is given by
\begin{eqnarray}
j = j_{R} + j_{L} + \int_{-\chi}^{\chi} \tilde{\rho}(\phi) \ d\phi
\end{eqnarray}
with
\begin{eqnarray}
\tilde{\rho}(\phi) - \int_{-\chi}^{\chi} K_{2}(\phi - \phi')\tilde{\rho}(\phi') \ d\phi' = j_{R}K_{2}(\chi - \phi) + j_{L}K_{2}(\chi + \phi) - \frac{1}{2\pi L}\sum_{\alpha = 1}^{J_{2}} \frac{1}{\frac{1}{4} + (v_{\alpha} - \phi)^2} \ \ 
\end{eqnarray}
and
\begin{eqnarray}
K_{2}(\phi) = \lim_{C_{g} \rightarrow 0} \frac{2\pi}{C_{g}} K\left(\frac{2\pi\phi}{C_{g}}\right).
\end{eqnarray}
Notice that the source term that depends on the $v_{\alpha}$ only survives the limit in the integral equation for $\tilde{\rho}(\phi)$, the non-movers; it does not directly affect the left- or right-movers.  We can then break up the equation for $\tilde{\rho}(\phi)$ the way we discussed in the previous section, so as to isolate the dependence on the $v_{\alpha}$s.  Noting that with the equation
\begin{eqnarray}
\tilde{\rho}_{pure}(\phi) - \int_{-\chi}^{\chi}K_{2}(\phi - \phi')\tilde{\rho}_{pure}(\phi') \ d\phi' = j_{R}K_{2}(\chi - \phi) + j_{L}K_{2}(\chi + \phi)
\end{eqnarray}
we will have
\begin{eqnarray}
\int_{-\chi}^{\chi} \tilde{\rho}_{pure}(\phi) \ d\phi = (1 - h(\chi))(j_{R} + j_{L})
\end{eqnarray}
for some function $h(\chi)$, we can finally write the relationship
\begin{eqnarray}
\sqrt{\frac{2h^2(\chi)\mathcal{E}L^2}{(1 - \tilde{K})\pi}} - J = \sum_{n = 0}^{\infty} \frac{g_{n}(\chi)}{2\pi}\left[\sum_{\alpha = 1}^{J_{2}} Q_{n}\left(\frac{1}{\frac{1}{4} + v_{\alpha}^2}\right)\right].
\end{eqnarray}
We know that the square root of the worldsheet energy density here should be proportional to the space-time energy density, and in turn to the dimension of the dual operator in the gauge theory.  We don't know what the function $h(\chi)$ is, since this depends on the S-matrix of the theory, which we don't have, but we know that this represents some dependence on the coupling $\lambda$.  Suppose we consider this equation in the case of a BPS state, where $J_{2} = 0$ so the right-hand-side of this equation vanishes.  In this case, we know we should have $\Delta = J$, so for more general states the left-hand-side of this expression must just be $\gamma$, the anomalous dimension!  The $g_{n}(\chi)$ are defined by
\begin{eqnarray}
g_{n}(\chi) = \int_{-\chi}^{\chi} f_{n}(\phi) \ d\phi
\end{eqnarray}
with
\begin{eqnarray}
f_{n}(\phi) - \int_{-\chi}^{\chi} K_{2}(\phi - \phi')f_{n}(\phi') \ d\phi' = \phi^{2n}.
\end{eqnarray}
They encode the information about the S-matrix of the theory in the expression for $\gamma$, while the polynomials $Q_{n}(x)$ include the information about the $SU(2)$ structure.  Comparing this expression with equation (\ref{eqn3}) from the loop expansion of the gauge theory, we find complete agreement in the structural dependence on the $v_{\alpha}$.  Here, the precise way that the conformal limit is taken conspires with the expansion of $\tilde{\rho}(\phi)$ to produce the expansion for $\gamma$ that matches the gauge theory result.  It was not obvious, before the conformal limit was taken, that the relationships between the energy density of a state, the number density $j$, and the dimension of the dual operator would produce exactly an anomalous dimension with a linear dependence on the polynomials $Q_{n}(x)$.  

Now, the equations (\ref{eqn4}) which quantize the $v_{\alpha}$s are not much affected by the conformal limit, except the $h_{n}(\lambda)$ are replaced by $h_{n}(\chi)$, defined by
\begin{eqnarray}
h_{n}(\chi) = \frac{(j_{R} + j_{L})\chi^{2n}}{j} + \frac{1}{j}\int_{-\chi}^{\chi} \phi^{2n}\tilde{\rho}(\phi) \ d\phi.
\end{eqnarray}
The idea behind the expansion (\ref{eqn4}) was to separate out the dependence on $u_{\alpha}$ from the dependence on the coupling, but it's not obvious from the above expression that this has been achieved.  We want to take the above expression and expand $\tilde{\rho}$ as before.  Making the definition
\begin{eqnarray}
r_{n}(\chi)(j_{R} + j_{L}) = \int_{-\chi}^{\chi} \phi^{2n}\rho_{pure}(\phi) \ d\phi
\end{eqnarray}
we end up with
\begin{eqnarray}
h_{n}(\chi) = \frac{\chi^{2n} + r_{n}(\chi)}{h(\chi)} - \frac{1}{j}\sum_{m = 0}^{\infty} \left[\frac{1}{2\pi L}\sum_{\alpha = 1}^{J_{2}} Q_{m}\left(\frac{1}{\frac{1}{4} + v_{\alpha}^2}\right)\right] \int_{-\chi}^{\chi} \left[\phi^{2n} - \frac{\chi^{2n} + r_{n}(\chi)}{h(\chi)}\right]f_{n}(\phi) \ d\phi \ \ 
\end{eqnarray}
Now, if the expansion (\ref{eqn4}) is to agree with the all loop guess, the terms in this equation that are dependent on the $v_{\alpha}$ must vanish.  We can see that it is possible that an S-matrix exists for which the integrals to vanish, but for general $f_{n}(\phi)$, they will not.  

\subsection{Expansion for the Hubbard Model}

The previous section demonstrates one way in which the $SU(2)$ structure for a nested Bethe ansatz could reproduce the all-loop guess, through a particular conformal limit.  But it turns out not to be the only way.  Recent work has shown that the all-loop guess for the $SU(2)$ sector of he gauge theory can be reproduced in a certain sector of the Hubbard model, a system of electrons on a one dimensional lattice \cite{Hubbard1}.   In this model, we say the lattice has $J$ sites and each site can be in one of four states: empty, occupied by one spin up electron, occupied by one spin down electron, or occupied by one of each (the Pauli exclusion principle forbids any other states).  This lattice system can be thought of as somewhere in between a spin chain and a continuous integrable field theory.  If we consider the sector of the theory that is ``half-filled'', where there are $J$ electrons on the lattice, then the state where all electrons are spin up is dual to the operator $\mbox{Tr} Z^{J}$.  $SU(2)$ impurities ($J_{2}$ of them) are then reproduced as spin down electrons in the lattice.  The Hubbard model is integrable, and the Hamiltonian in this sector is diagonalized using a nested Bethe ansatz, just as we expect in the worldsheet theory:
\begin{eqnarray}
\label{eqn5}
e^{iq_{i}J} = \prod_{\alpha = 1}^{J_{2}} \frac{u_{\alpha} - \sqrt{2}g\sin q_{i} - i/2}{u_{\alpha} - \sqrt{2}g\sin q_{i} + i/2}, \ \ \ \ \ \ i = 1, ..., J
\end{eqnarray}
\begin{eqnarray}
\label{eqn6}
e^{ip_{\alpha}J} \equiv \prod_{i = 1}^{J} \frac{u_{\alpha} - \sqrt{2}g\sin q_{i} - i/2}{u_{\alpha} - \sqrt{2}g\sin q_{i} + i/2} = \prod_{\beta \ne \alpha}^{J_{2}} \frac{u_{\alpha} - u_{\beta} + i}{u_{\alpha} - u_{\beta} - i}, \ \ \ \ \ \ \alpha = 1, ..., J_{2}
\end{eqnarray}
\begin{eqnarray}
E = \frac{\sqrt{2}}{g}\sum_{i = 1}^{J} \cos q_{i}.
\end{eqnarray}
I am assuming that the number of sites $J$ is odd; if it is not there are additional ``twisting'' factors.  In this nesting structure, the first Bethe ansatz represents the population of the $J$ sites with $J$ electrons, and the second represents spin down impurities traveling around the spin chain that now exists on these sites.  This is very much like what we expect for the worldsheet.  We can see that the $SU(2)$ structure presents itself in the same way as before.  One major difference is the lack of terms associated with the interactions of two spin up electrons: in the Hubbard model, these electrons do not interact.  Another is the explicit presence of dependence on a coupling $g$, which is related to $\lambda$ by $g^2 = \frac{\lambda}{8\pi^2}$.  

We know that this system agrees with the all-loop guess in the large $J$ limit (where the all-loop guess is believed to be valid).  In this limit we replace individual $q_{i}$ with a density in momentum space $\rho(q)$ so that equation (\ref{eqn5}) becomes
\begin{eqnarray}
\rho(q) = \frac{1}{2\pi} + \frac{\sqrt{2}g\cos q}{2\pi L}\sum_{n = 0}^{\infty} \left(2g^2\sin^2 q\right)^{n} \sum_{\alpha = 1}^{J_{2}} Q_{n}\left(\frac{1}{\frac{1}{4} + u_{\alpha}^2}\right)
\end{eqnarray}
where we are again neglecting the odd part of $\rho(q)$ because it will not contribute to to energy of the state, and expanding in the usual way.  Here, expansion in small $g$ is the same as a ``relativistic'' expansion which would assume $\sin q_{i} < u_{\alpha}$ for all $i$ and $\alpha$.  The half-filling condition becomes
\begin{eqnarray}
1 = \int_{-B}^{B} \rho(p) \ dp
\end{eqnarray}
which fixes the limits of integration to be $B = \pi$.  The energy of the state is then 
\begin{eqnarray}
E = \frac{\sqrt{2}L}{g}\int_{-\pi}^{\pi} \cos q \ \rho(q) \ dq = \sum_{n = 0}^{\infty} \frac{(2n)!}{(n+1)!n!}\left(\frac{\lambda}{16\pi^2}\right)^{n}\left[\sum_{\alpha = 1}^{J_{2}} Q_{n}\left(\frac{1}{\frac{1}{4} + u_{\alpha}^2}\right)\right].
\end{eqnarray}
which reproduces equation (\ref{eqn3}) up to an overall constant.  Equation (\ref{eqn6}) becomes
\begin{eqnarray}
\frac{p_{\alpha}}{2} = \sin^{-1}\sqrt{\frac{1}{1 + 4u_{\alpha}^2}} + \frac{u_{\alpha}}{2}\sum_{n = 0}^{\infty}h_{n+1}(\lambda)P_{n}\left(\frac{1}{\frac{1}{4} + u_{\alpha}^2}\right)
\end{eqnarray}
just as before, and here
\begin{eqnarray}
h_{n}(\lambda) = \int_{-\pi}^{\pi} 2^{n}g^{2n}\sin^{2n}q \ \rho(q) \ dq.
\end{eqnarray}
Again, we need to insure that the $h_{n}(\lambda)$ are truly independent of the $u_{\alpha}$; in this case it is acheived because
\begin{eqnarray}
\int_{-\pi}^{\pi} \sin^{2\ell} q \ \cos q \ dq = 0
\end{eqnarray}
for any integer $\ell$, and finally we have
 \begin{eqnarray}
\frac{p_{\alpha}}{2} =  \sin^{-1}\sqrt{\frac{1}{1 + 4u_{\alpha}^2}} + u_{\alpha}\sum_{n = 0}^{\infty} \left(\frac{\lambda}{16\pi^2}\right)^{n+1}\left(\begin{array}{c} 2n + 1 \\ n \end{array}\right) P_{n}\left(\frac{1}{\frac{1}{4} + u_{\alpha}^2}\right)
\end{eqnarray}
which agrees with equation (\ref{eqn7}).  This loop by loop agreement is no surprise, given that the exact relationship has already been demonstrated.  What is interesting is how much of this agreement can be shown to arise from the $SU(2)$ structure alone.  Now, the Hubbard model is a lattice system and is not clearly related to the worldsheet theory, but it's nested Bethe ansatz structure is obviously similar to what we expect for the worldsheet theory, and it produces the necessary functions $h_{n}(\lambda)$ and $g_{n}(\lambda)$ to complete the duality at each loop level.  

The question now becomes just how difficult this is to do.  We know the values of the $g_{n}(\lambda)$ and $h_{n}(\lambda)$ are determined by the S-matrix for identical particles in the theory and also by the way a conformal limit is taken and the dependence on a coupling constant arises.  It is possible that there is essentially only one way the correct $g_{n}(\lambda)$ and $h_{n}(\lambda)$ can be obtained, so that the values of these essentially determine the S-matrix, the conformal limit, and the dependence on the coupling.  If this is true then the equations (\ref{eqn5}) and $(\ref{eqn6})$ must essentially be the right answer for the worldsheet theory and must have some interpretation in these terms as a continuous theory rather than a lattice theory.

\section{Conclusions}

Here we have shown that the structure of an $SU(2)$ nested Bethe ansatz in a continuous integrable system can be directly related to the all-loop guess for an $SU(2)$ sector of operators in $\mathcal{N} = 4$ super Yang-Mills theory.  Without knowing much about the theory beyond that it is integrable and possesses an $SU(2)$ sector we can reproduce many of the features of this all-loop guess and we can isolate those aspects of the Bethe ansatz that depend on the details of the theory from those that depend only on the $SU(2)$ symmetry.  This is done by identifying a relativistic expansion for the worldsheet theory with the loop expansion of the gauge theory.  Finally, we have discussed how this expansion plays out in the known relationship between the nested Bethe ansatz structure of a sector of the Hubbard model and the all-loop guess.

One interesting question that this research brings up is whether or not this way of looking at the all-loop guess and perhaps the Hubbard model can be ``reverse engineered'' to yield information about the worldsheet theory.  This would depend on how restrictive matching the details of the all-loop guess is; whether or not this matching really includes all information about the S-matrix for identical particle interactions in the worldsheet theory, for example.  

Another issue is what this might tell us about the validity of the all-loop guess, which was derived by assuming both integrability and BMN scaling are exact; without assuming BMN scaling there are other long range integrable spin chains that could be identified with the dilatation operator for single trace $SU(2)$ operators.  For a continuous integrable theory with a nested Bethe ansatz, the $SU(2)$ structure and the assumption of integrability alone reproduce much of the all-loop guess, it would be interesting to know if they are consistent with any of the long range spin chains that don't assume BMN scaling.  It's possible that some or all of these other options would not be consistent, which might be seen as indirect evidence in favor of BMN scaling.

Finally, it's possible that identifying the relativistic expansion in a worldsheet nested Bethe ansatz with the loop expansion in the gauge theory might give us some clues as to how finite size effects affect the worldsheet theory, for previous work on finite size corrections see \cite{finite}.  We know that finite size effects arise in the gauge theory in a very discrete manner: the one loop term is correct for all the single-trace operators, but as we add loop corrections we restrict the guess to longer and longer operators: at $L$ loops the guess breaks down for operators of length $L$ and shorter.  If loop corrections in the gauge theory are dual to relativistic corrections in the worldsheet theory, then this implies that these relativistic corrections are also related to finite size effects.  Perhaps relativistic corrections gradually de-localize interactions between particles in the theory and thus the ansatz's validity at a particular order in relativistic corrections is related to the size of the worldsheet spatial dimension.

\section*{Acknowledgments}

I would like to thank J. Polchinski, N. Beisert, T. Erler, R. Roiban, M. Staudacher, and F. Cachazo for useful discussions, and J. Wong for her hospitality during some of the research.  This work was supported by National Science Foundation grant PHY00-98395.  Any opinions, findings, and conclusions or recommendations expressed in  this material are those of the author and do not necessarily reflect the views of the National Science Foundation.

\end{document}